# PHASE COMPETITION IN $Ln_{0.5}A_{0.5}MnO_3$ PEROVSKITES

F. Rivadulla, E. Winkler, J.-S. Zhou, J. B. Goodenough

*Texas Materials Institute, , ETC 9.102,The University of Texas at Austin, Austin, TX 78712-1063 (USA)*

**ABSTRACT**

Single crystals of the systems $Pr_{0.5}(Ca_{1-x}Sr_x)_{0.5}MnO_3$, $(Pr_{1-y}Y_y)_{0.5}(Ca_{1-x}Sr_x)_{0.5}MnO_3$, and $Sm_{0.5}Sr_{0.5}MnO_3$ were grown to provide a series of samples with fixed ratio Mn(III)/Mn(IV)=1 having geometric tolerance factors that span the transition from localized to itinerant electronic behavior of the $MnO_3$ array. A unique ferromagnetic phase appears at the critical tolerance factor $t_c= 0.975$ that separates charge ordering and localized-electron behavior for $t<t_c$ from itinerant or molecular-orbital behavior for $t>t_c$. This ferromagnetic phase, which has to be distinguished from the ferromagnetic metallic phase stabilized at tolerance factors $t>t_c$, separates two distinguishable Type-CE antiferromagnetic phases that are metamagnetic. Measurements of the transport properties under hydrostatic pressure were carried out on a compositions t a little below $t_c$ in order to compare the effects of chemical vs. hydrostatic pressure on the phases that compete with one another near $t=t_c$.

## I.- INTRODUCTION

In the $Ln_{1-x}A_xMnO_3$ perovskites, a transition from localized to itinerant electronic behavior can be induced either by hole doping or by changing the tolerance factor t at a given $x\geq0.15$.[1] The transition is first-order, and when phase segregation occurs at too low a temperature for atomic diffusion, it may be accomplished in a perovskite by cooperative oxide-ion displacements.[2] As occurs with a Jahn-Teller deformation, these cooperative displacements may be long-range ordered and static or short-range ordered in a fluctuating spinodal phase segregation that occurs at a small length scale.[3] At x=0.5, the Coulombic repulsion between localized electrons induces charge ordering (CO) below a certain temperature, leading to a rich variety of charge-ordered/orbital-ordered (CO/OO) structures depending on the compostion.[4,5] On the other hand it is also necessary to consider the lattice instabilities and competing phases that enter at a crossover from localized to itinerant electronic behavior. For large values of the tolerance factor, as in $La_{0.5}Sr_{0.5}MnO_3$, the material behaves like a metallic ferromagnet (FMM) down to the lowest temperature, demonstrating the key role of the tolerance factor in the stabilization of a broad σ* band in which de Gennes double exchange couples the spins ferromagnetically.[6]

Kuwahara and Tokura[4] presented a tentative phase diagram for half doped manganites (see figure 1); by comparison with their data at x=0.45 a phase diagram resembling quantum-critical-point (QCP) behavior was proposed to occur at a critical tolerance factor $t_c$ in the $Ln_{0.5}A_{0.5}MnO_3$ perovskites.[7] At the QCP, quenched disorder is introduced in the Hamiltonian describing the competition between FM and CO ground states. Tokura *et al.*[8] also reported for the first time the existence of a very narrow FMM region inserted between two antiferromagnetic CE (CO/OO) phases. However, this phase was considered to be a metastable state and consequently its relevance for understanding the phase diagram was not properly highlighted.

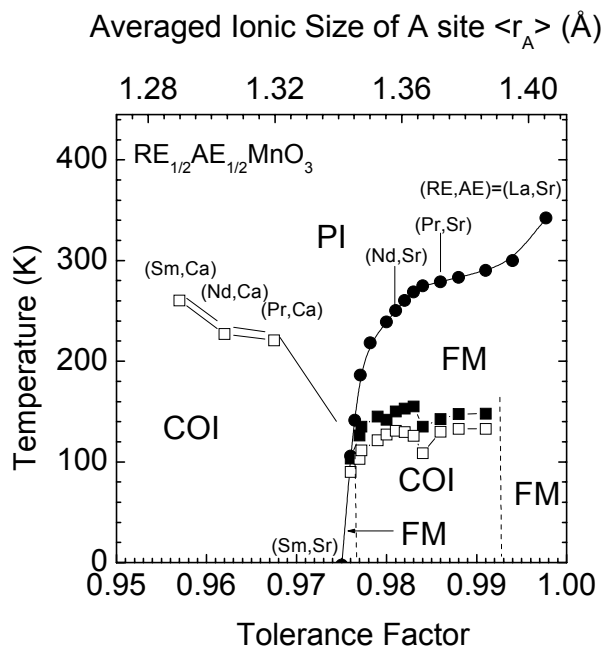

**Figure 1.-** Phase diagram for $Ln_{0.5}A_{0.5}MnO_3$ proposed by Kuwahara and Tokura in reference [4]. In this diagram, PI= paramagnetic insulating, COI= charge ordering insulating, FM= ferromagnetic metallic.

On the other hand, Rao *et al.*[5] proposed a schematic phase diagram for the half-doped systems in which only the main magnetic phases were considered; it provides no proper description of the regions between the well-established magnetic phases in which two or more phases compete. Moreover, the FMM phase reported by Tokura does not appear in this oversimplified phase diagram. In another phase diagram, presented by Damay *et al.*,[9] neither the FMM phase nor the multiple-phase regions are considered; in this case, the authors were concerned to report a detailed description of the evolution of the crystallographic transitions with the tolerance factor, but a conventional diffraction experiment is unable to detect two-phase fluctuations.

In this paper, we examine in detail the relative stabilities of localized and itinerant-electron phases as the tolerance factor is varied from $t<t_c$ to $t>t_c$. We demonstrate the existence of a unique ferromagnetic phase appearing at a crossover of a CO temperature $T_{CO}$ and the Curie temperature $T_C$ for a FMM phase at a critical tolerance factor $t=t_c$. The unique FMM phase appearing at the crossover has a larger volume in the paramagnetic state at room temperature than expected from the evolution of volume with t and is therefore suppressed by pressure; it separates two Type-CE antiferromagnetic phases we propose need to be distinguished. We predict that in one the electrons are localized with CO of Mn(III) and Mn(IV) ions as originally predicted for the CE phase;[10] in the other we anticipate an ordering of Zener two-manganese Mn(III)-O-Mn(IV) pairs containing a molecular orbital within a pair.

We also identify the stabilization of superparamagnetic (SP) clusters within a paramagnetic (PM) matrix with CO fluctuations below a $T^*\approx 380$ K$>T_{CO}$ or $T_C$ over a range of tolerance factors $t_c\pm\Delta t$. We summarize our findings with that of others in the phase diagram at the end of this paper (see figure 15).

## II.- EXPERIMENTAL

Our studies were performed on single crystals of the systems $Pr_{0.5}(Ca_{1-x}Sr_x)_{0.5}MnO_3$, $(Pr_{1-y}Y_y)_{0.5}(Ca_{1-x}Sr_x)_{0.5}MnO_3$, and $Sm_{0.5}Sr_{0.5}MnO_3$ in order to vary t across $t_c$ for a fixed ratio Mn(III)/Mn(IV)=1. All the crystals were grown by the floating-zone technique in an IR radiation image furnace under flowing $O_2$. Stoichiometric proportions of the starting materials $Pr_6O_{11}$, $CaCO_3$, $SrCO_3$, $Mn_2O_3$, $Sm_2O_3$ and $Y_2O_3$ were homogenized, pressed into a pellet, and fired at 1100ºC for 24 h before grinding again. This procedure was repeated three or four times for each composition. The powder was pressed hydrostatically into a cylindrical shape to make a feed rod. The crystals were grown from the rods at a rate of 3 mm/h. Laue back diffraction and powder diffraction were used to verify the single-crystal character and structure of the samples; the peaks were fully indexed in space group *Pbnm*. We calculate t on the basis of 12-fold coordination of the A-site cations. Fig. 2 compares the variation with t of the room-temperature volume of the unit cell of our crystals with those obtained by Kuwahara *et al.*[11] with the system $(Nd_{1-x}Sm_x)_{0.5}Sr_{0.5}MnO_3$. The volume increases progressively with the mean size of the larger cations except near the composition with $t\approx t_c$ where there is an anomalous increase in the room-temperature volume followed, with increasing t, by an abrupt drop. We define $t=t_c$ as the tolerance factor where a unique ferromagnetic metallic phase appears. This peculiar FMM phase is stable in a very narrow range of t and stoichiometry. We tried to discard any deviation of the stoichiometry as the source of this anomaly in the volume. Iodimetric analysis of the crystals gave %$Mn^{4+}$=50.5(5) in all the cases. In the ceramic precursors used to grow the crystals, the result was 50.0% within the error of the measurement. This small difference between ceramics and single-crystal analysis could be due to the evaporation of a small amount of the trivalent ion ($La^{3+}$ vacancies) during crystal growth or, more probable, to an error in the analytical determination due to difficulties to achieve complete solubilization of the crystalline material in diluted HCl. We conclude that $Mn^{3+}/Mn^{4+}\approx 1$ in our samples and that the large volume at $t_c$ cannot be due to a reduction in the amount of $Mn^{4+}$.

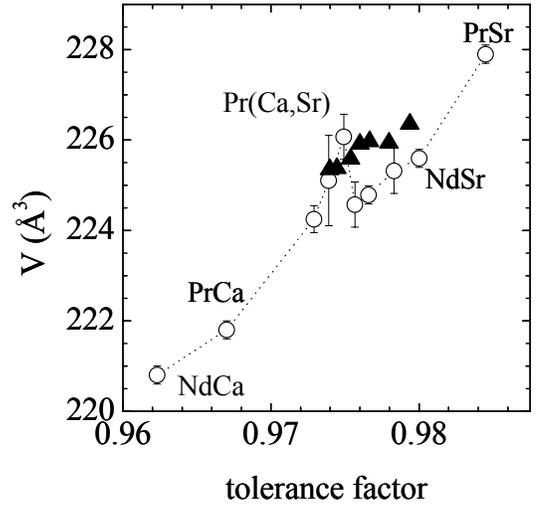

**Figure 2.-** Evolution of the room temperature volume with tolerance factor. Filled triangles are taken from ref. [11] for the series $(Nd_{1-x}Sm_x)_{0.5}Sr_{0.5}MnO_3$.

Our crystals of the series $Pr_{0.5}(Ca_{1-x}Sr_x)_{0.5}MnO_3$ have a small variance $\sigma^2=\Sigma x_i r_i^2-<r>^2$ of the A-site sizes; $x_i$ is the concentration of each A-site cation of ionic radius $r_i$ and $<r>$ is the mean radius of the A-site cations.[12] Other systems with larger $\sigma^2$, like $(Pr_{1-y}Y_y)_{0.5}(Ca_{1-x}Sr_x)_{0.5}MnO_3$, were synthesized to study the effect of A-site disorder on the critical temperatures around $t_c$. In these systems, as in $(Nd_{1-x}Sm_x)_{0.5}Sr_{0.5}MnO_3$, the $T_C$ of the FM phase was significantly lower, indicating a marked influence of the variance on the transition temperatures around $t_c$.

A four-probe method was used to measure the resistivity. The thermoelectric power $\alpha(T)$ and resistivity $\rho(T)$ measurements under pressure were carried out in a self-clamped Be-Cu cell with silicone oil as the pressure medium. The pressures indicated in this paper correspond to those at *ca.* 225 K, which is near the charge-ordering temperature $T_{CO}$ in $Pr_{0.5}Ca_{0.5}MnO_3$. Magnetization was measured with a SQUID magnetometer (Quantum Design). Thermal conductivity was measured with a steady-state method; the temperature gradient was controlled to be less than 1% of the base temperature.

## III.- RESULTS AND DISCUSSION

The high-temperature paramagnetic susceptibilities of all our samples, *e.g.* Fig. 3, shows Curie-Weiss behavior above a temperature $T^* \approx 380 \pm 10$ K with a positive Weiss temperature $\theta > 0$ characteristic of ferromagnetic Mn-O-Mn interactions; the rare-earth interactions are much weaker. At these temperatures the e electrons are localized and the occupied e orbitals fluctuate; the ferromagnetic interactions reflect a combination of vibronic superexchange and a Zener double exchange associated with mobile two-manganese Zener polarons. A deviation from linearity of $\chi^{-1}(T)$ below $T^*$ is characteristic of either short-range order or the onset of superparamagnetic (SP) clusters.[13] Given a $T^* \gg \theta$, the formation of SP-Zener polarons is the more likely cause, and this deduction is supported by the thermoelectric-power data (see discussion of Eq. (1)). Therefore, Fig. 1 identifies SP clusters in a paramagnetic (PM) matrix in the temperature interval $T_{CO}$, $T_C < T < T^*$. It is only below the transition at $T_{CO}$ or $T_C$ that the system distinguishes localized electrons for a tolerance factor $t < t - \Delta t_1$ from itinerant or molecular orbital electrons for $t > t_c + \Delta t_2$, where $t_c \approx 0.975$.

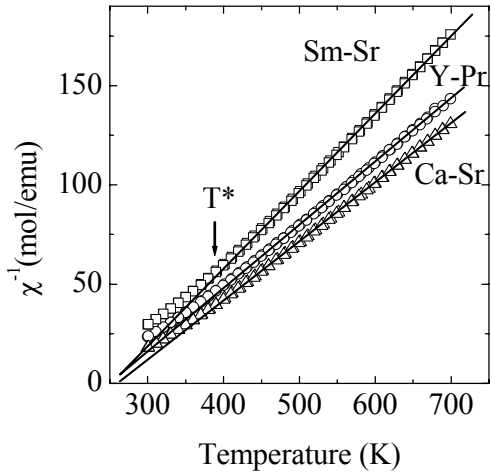

**Figure 3.-** Temperature dependence of the inverse susceptibility (H=1 T) in the paramagnetic range for some representative crystals. The line is the fit to the Curie-Weiss law. Sm-Sr, Y-Pr, and Ca-Sr, represent $Sm_{0.5}Sr_{0.5}MnO_3$, $(Pr_{0.9}Y_{0.1})_{0.5}(Ca_{0.655}Sr_{0.345})_{0.5}MnO_3$, and $Pr_{0.5}(Ca_{0.85}Sr_{0.15})_{0.5}MnO_3$ respectively.

The compound $Pr_{0.5}(Ca_{0.9}Sr_{0.1})_{0.5}MnO_3$ (t=0.973) appears to lie just within the two-phase interval $\Delta t_1$ in our phase diagram (see Fig. 15). The $\rho(T)$ data, inset of Fig. 4(a), show a first-order phase change at a charge-ordering temperature $T_{CO} \approx 225$ K. The magnetization M(T) in a field of 10 Oe can be seen in Fig. 4(a) to exhibit a sharp maximum at $T_{CO}$ and a smaller maximum at $T_N=150$ K below which Type-CE antiferromagnetic order has been found;[14] the magnetic Type-CE order indicates that long-range orbital ordering has also occurred below a $T_{OO} \approx T_N$.

For the sake of simplicity, in the following discussion and in the phase diagram that we present at the end of this paper, we will refer to $T_{OO}$ as a long-range orbital-ordering temperature. That is why we identify $T_{OO} = T_N$, although it is necessary to have in mind that this is not absolutely true when $T_{CO} > T_N$, where the continuous structural deformation when the temperature is reduced show that the OO is continuously developed above $T_N$.[15]

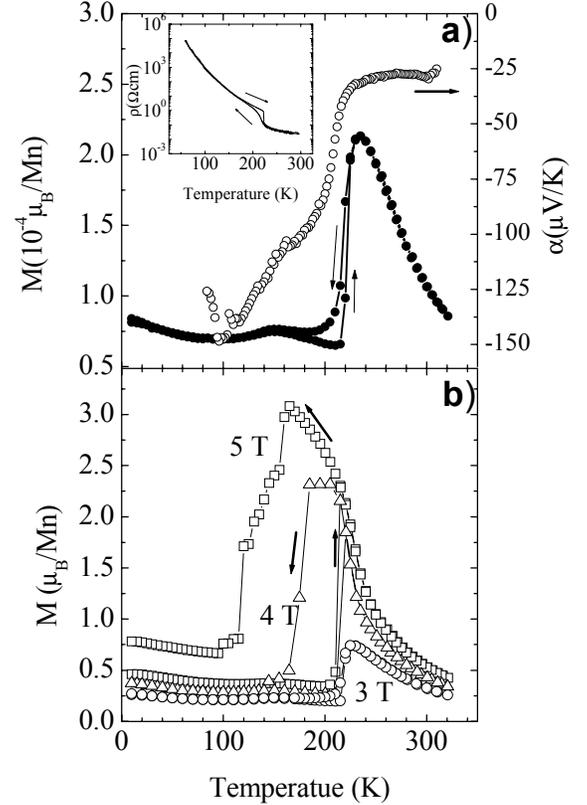

**Figure 4.-** (a) Temperature dependence of the magnetization (FC-ZFC, H=10 Oe) and thermoelectric power for $Pr_{0.5}(Ca_{1-x}Sr_x)_{0.5}MnO_3$, x=0.1; t=0.973. Inset: Resistivity vs temperature on cooling and warming. (b) Magnetization at several fields. For H=5 T the sample almost reaches saturation (3.5 $\mu_B$) between $T_C$ and $T_N$.

The thermoelectric power $\alpha(T)$ of Fig. 4(a) sheds additional light on what is happening. A temperature-independent $\alpha$ at $T > T_{CO}$ is typical of polaronic conduction dominated by the statistical contribution

$$\alpha = (k_B/e)\ln[\beta(1-Qc)/Qc] \qquad (1)$$

The spin-degeneracy factor is $\beta=1$ because of the strong intraatomic exchange, and a ratio Mn(III)/Mn(IV)=1 makes c=0.5. For small polarons, the number of sites occupied by a charge carrier is Q=1, which would give $\alpha \approx 0$, whereas we observe an $\alpha = -28$ $\mu$V/K requiring a $Q \approx 1.2$. Zener polarons would have a Q=2. Therefore, the data suggest that about 20% of the e electrons have become delocalized within two-manganese molecular orbitals, which is consistent with our conclusions from the magnetic-susceptibility data. Moreover, similar values of $\alpha$ for $T>T_{CO}$ or $T_C$ have been found by other authors[16] for various compositions with x=0.5. This conclusion is also consistent with the observation of a crossover with increasing t at lower temperatures from localized e electrons to itinerant electrons in a narrow $\sigma^*$ band of e-orbital parentage. The formation of two-manganese Zener polarons represents a first step in the transition to itinerant behavior. Finally, this interpretation allows us to understand the sharp drop in M(T) on cooling through $T_{CO}$ as a breakup of the Zener polarons by the charge ordering, which localizes the holes at ordered Mn(IV) sites. The charge ordering traps out mobile holes, so the magnitude of $\alpha(T)$ increases sharply on cooling through $T_{CO}$, and a further increase on cooling through $T_N$ indicates a long-range $T_{OO} \approx T_N$ in this material. Independent neutron-diffraction experiments[15] have revealed that short-range orbital ordering increases on cooling in the interval $T_N<T<T_{CO}$. The CO and short-range orbital order appear to introduce a frustration that suppresses long-range magnetic order, which is why long-range $T_{OO} \approx T_N$.

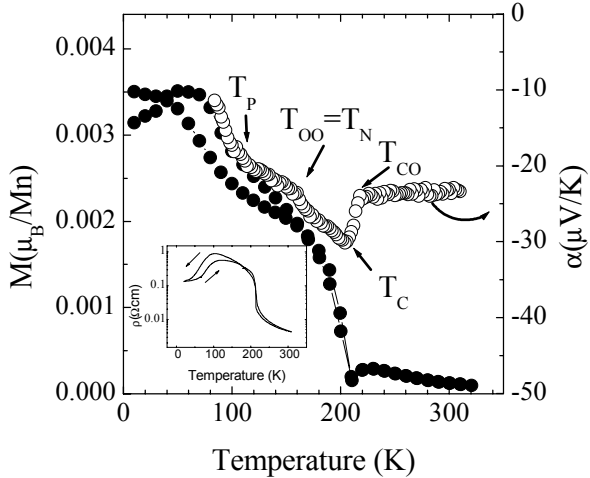

**Figure 5.-** Temperature dependence of the magnetization (FC-ZFC, H=10 Oe) and thermoelectric power for $Pr_{0.5}(Ca_{1-x}Sr_x)_{0.5}MnO_3$, x=0.15; t=0.974. Inset: Resistivity vs temperature on cooling and warming.

Fig. 4(b) shows, for $Pr_{0.5}(Ca_{0.9}Sr_{0.1})_{0.5}MnO_3$, the change in M(T) in an applied magnetic field in the interval $3\ T \leq H \leq 5\ T$. In an H= 4.5 T, static charge ordering is completely suppressed in the range $T_N<T<T_{CO}$; a ferromagnetic state is stabilized below a $T_C$ = 227 K that saturates at 3.2 $\mu_B$/Mn above $T_N$, near the spin-only 3.5 $\mu_B$/Mn. Orbital ordering below $T_{OO} \approx T_N$ reestablishes the static charge ordering that is suppressed by the ferromagnetic phase. Stabilization by a magnetic field of an orbitally disordered ferromagnetic phase relative to an orbitally ordered antiferromagnetic phase has been demonstrated[17] in the systems $LaMn_{1-x}Ga_xO_3$ and $LaMn_{1-x}Sc_xO_3$ whereas in this mixed-valent phase the applied field stabilizes a ferromagnetic phase with either charge disorder or suppressed charge separation relative to a charge-ordered phase that breaks up the Zener polarons. A similar behavior has been found by Kuwahara and Tokura in $Pr_{0.5}Ca_{0.5}MnO_3$, but in higher magnetic fields as a consequence of a smaller tolerance factor t.[4]

In the interval 0.973<t<0.978 (see Figs. 1 and 15) shows the appearance of two phases; a unique ferromagnetic (FMM) phase appears at $t_c$=0.975 in which OO and CO are suppressed. This FMM phase appears to be distinguishable from the FM phase stabilized in an H≥4.5 T at t=0.973; it appears as a single phase where $T_C$ at H=0 crosses $T_{CO}$ at H=0 in the phase diagram.

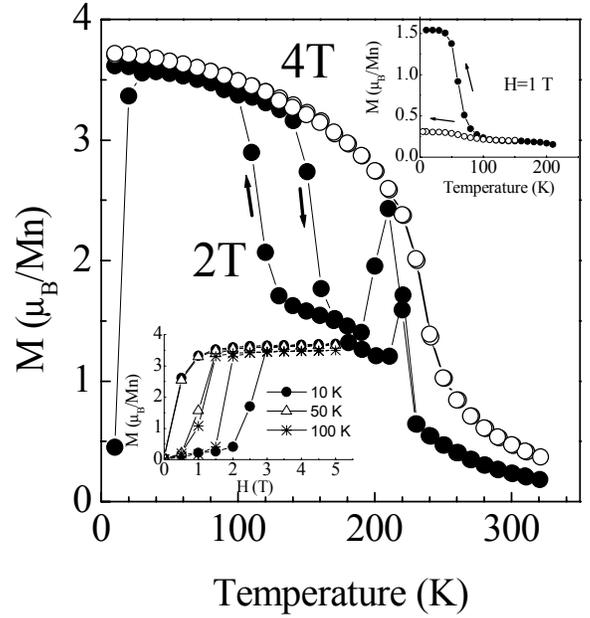

**Figure 6.-** Effect of an applied magnetic field on the magnetization of $Pr_{0.5}(Ca_{1-x}Sr_x)_{0.5}MnO_3$, x=0.15. Upper inset: The open circles curve was measured in FC (H=1T) after ZFC from 320 down to 5 K and warming up to 150 K to in zero field. The curve with solid circles was measured in the same way but warming up to 210 K. Lower inset: Magnetization vs. field at different temperatures below $T_P$.

A $Pr_{0.5}(Ca_{0.85}Sr_{0.15})_{0.5}MnO_3$ crystal with t=0.974 probes the two-phase region below $T_{CO}$ in the interval $\Delta t_1$ between t=0.973 and $t_c$=0.975. The $\rho(T)$ curve of the inset of Fig. 5 shows a $T_{CO} \approx 220$ K. However, unlike the t=0.973 sample of Fig. 4(a), the M(T) curve for t=0.974 shows a ferromagnetic minority phase in a field H=10 Oe appearing below a $T_C$=210 K. The volume fraction of this phase apparently grows as the temperature is lowered, reaching a percolation threshold at a $T_p \approx 100$ K below which M(T) increases more sharply with decreasing temperature. The $\rho(T)$ curve shows a change to a lower resistivity below $T_p$, which is consistent with percolation of a more conductive ferromagnetic phase below this temperature. Fig. 6 shows that the ferromagnetic, vibronic FMV phase is more easily stabilized by a magnetic field relative to the antiferromagnetic (AF) phase at temperatures $T<T_P$. The $\alpha(T)$ curve increases in magnitude on cooling through $T_{CO}$, but the change is smaller than that at $T_{CO}$ in Fig.

4(a) because CO occurs in a smaller fraction of the volume. The α(T) curve also shows a decrease in magnitude with decreasing temperature below $T_C$ and the growth of its volume fraction to beyond percolation below $T_p$. A $T_{OO}=T_N\approx150$ K in the CO matrix appears as a change of slope in the α(T) curve. The thermal hysteresis in M(T) and ρ(T) below 150 K also signals a $T_{OO}=T_N\approx150$ K for the CO matrix.

Application of a magnetic field H=4 T to the t=0.974 sample completely suppresses the CO phase, Fig. 6. A field of H=2 T raises $T_C$ above $T_{CO}$ and suppresses the AF phase below $T_{OO}=T_N$, but it does not suppress the CO phase in the paramagnetic temperature range of the CO phase. The AF phase is thus seen to be metamagnetic with the AF-FM transition occurring at $T_p$ on cooling in H=2 T whereas the FM phase is retained after zero-field cooling (ZFC) on heating in H=2 T to the magnetic transition temperature at $T_{OO}=T_N$ in zero field.

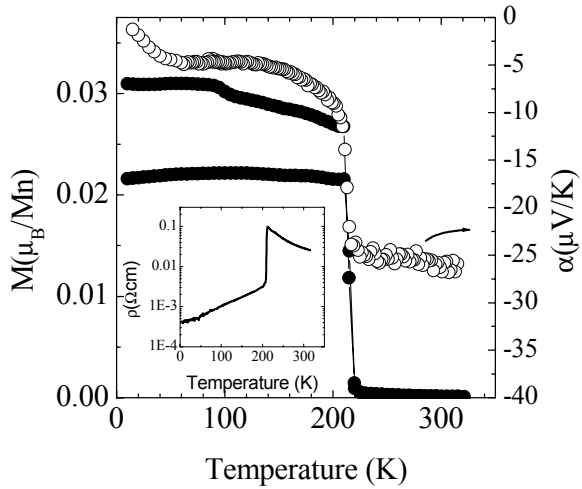

**Figure 7.-** Temperature dependence of the magnetization (●) FC-ZFC, H=10 Oe, and thermoelectric power (O) for $Pr_{0.5}(Ca_{1-x}Sr_x)_{0.5}MnO_3$, x=0.20; t=0.975. Inset: Resistivity vs temperature showing the abrupt metal insulator transition at ≈215 K.

The upper inset of Fig. 6 shows two M(T) curves taken on cooling in a field H=1 T after a rapid ZFC to 5 K and warming to 150 K and to 210 K. After warming to 150 K before H=1 T is applied, the field-cooled (FC) magnetization remains very small (open circles); the crystal remains in the AF state. However, after warming to 210 K before H=1 T is applied, the FC M(T) curve (closed circles) increases sharply on cooling below $T_p$. The metamagnetism of the AF phase at H=1 T is thus seen to require nucleation within it of a FM phase that grows below $T_p$. Rapid cooling to 5 K in zero field apparently does not provide time for the ferromagnetic clusters to nucleate. Uehara and Cheong have already shown how cooling rate and aging in a magnetic field can drastically influence the ratio of FM/CO phases that coexist in a wide temperature interval in the system $La_{5/8-y}Pr_yCa_{3/8}MnO_3$.[18]

The lower inset of Fig. 6 shows M vs H hysteresis for 0≤H≤5 T at three different temperatures after a ZFC. At 10 K, the metamagnetic transition occurs in the interval 2 T≤H≤3 T on increasing H; it returns to a spin-glass or canted-spin AF state on decreasing H below 1 T. At 50 K, the metamagnetic transition occurs in the interval 0.5 T<H<1.5 T in conformity with the upper inset. At 100 K ≈ $T_p$, the metamagnetic transition occurs at a larger magnetic field, 1.5 T<H< 1.8 T on raising H and returning to a spin-glass or canted-spin AF state at about H=1 T on reducing H. This remarkable shift confirms that growth of the FM phase at the expense of the AF phase is greatly facilitated below $T_p$.

Fig. 7 shows M(T) and M(H) curves for $Pr_{0.5}(Ca_{0.8}Sr_{0.2})_{0.5}MnO_3$ with t=0.975≈$t_c$. The abrupt appearance of ferromagnetism below $T_C$=220 K in H=10 Oe, the shift in $T_C$ to near 250 K in H=5 T, and the hysteresis in the M(H) curve at 210 K all indicate a $T_C\approx T_{CO}$=220 K in zero field with complete suppression of the CO phase in a field H≥1 T or at lower temperatures. The existence of this ferromagnetic phase is very elusive and has not been reported in many of the diagrams presented for half-doped manganites. For example, Damay et al.[9] presented a phase diagram for the series $Pr_{0.5}Sr_{0.5-x}Ca_xMnO_3$, but they missed this phase because they concentrated more on the low x part of the diagram and did not synthesize samples between x=0.5 and x=0.2. Similarly, this phase is not reported in other diagrams of half-doped manganites (see, for example, the one by Rao and coworkers in ref. [5]).

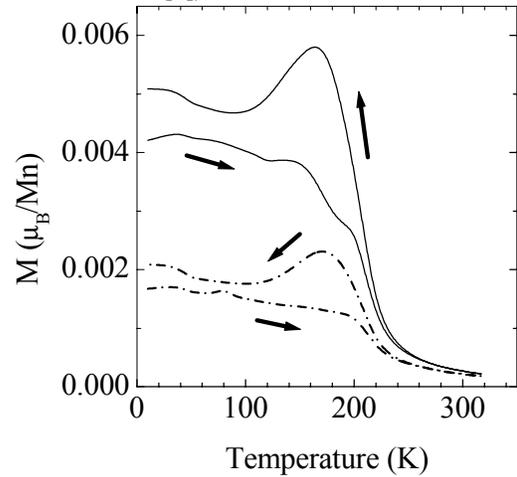

**Figure 8.-** Magnetization (H=10 Oe) vs temperature in $Pr_{0.52}(Ca_{0.8}Sr_{0.2})_{0.48}MnO_3$ (solid line) and $Pr_{0.48}(Ca_{0.8}Sr_{0.2})_{0.52}MnO_3$ (dashed line). The volume of the ferromagnetic phase in $Pr_{0.5}(Ca_{0.8}Sr_{0.2})_{0.5}MnO_3$ phase is destroyed by the small variation in the stoichiometry.

A much more complete characterization of the $Pr_{0.5}Sr_{0.5-x}Ca_xMnO_3$ system was made by Krupicka et al.[19], although they again failed to describe the ferromagnetic metallic phase at x=0.1. However, from their phase diagram, this phase cannot be discarded because these authors introduce a boundary between regions I and II of their diagram at what we refer to as $t_c$, but they did not examine a composition at this boundary. Their boundary separates a homogeneous Type-CE magnetic order from a two-phase CE+FMM mixture. Moreover, the authors found a peak in the susceptibility around 40 K that they attribute to the presence of $Mn_3O_4$ in their samples. We have demonstrated the narrow compositional range of stability of this unique FMM phase by varying the Mn(III)/Mn(IV) ratio from unity to $1\pm\delta$ in $Pr_{0.5\pm\delta}(Ca_{1-x}Sr_x)_{0.5\pm\delta}MnO_3$. The FM phase was suppressed for values of $\delta\geq 0.02$, Fig. 8.

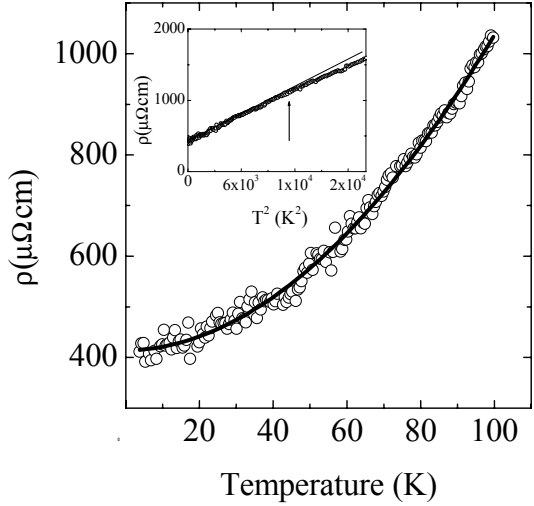

**Figure 9.-** The temperature dependence of the resistivity at low temperatures for $Pr_{0.5}(Ca_{1-x}Sr_x)_{0.5}MnO_3$, x=0.20. The solid line is the fit to Equation (3). Inset: The temperature dependence of the resistivity vs $T^2$ showing deviation at ≈100 K.

On the other hand, Tokura et al.[8] have reported the existence of a peculiar ferromagnetic phase sandwiched between two Type-CE AF phases at a $t=t_c$; however, their composition, $(Nd_{0.125}Sm_{0.875})_{0.5}Sr_{0.5}MnO_3$, has a larger variance $\sigma^2$, and the $T_C$ was consequently about 100 K lower than that we report here. Also Kiruki et al.[20] proved the existence of the unique FMM phase, and gave a tentative explanation for the anomalous response of its transport properties to hydrostatic pressure. As we will show in the rest of this paper, consideration of this narrow ferromagnetic region at t=0.975 is a necessary condition to understand correctly the phase diagram of half-doped manganites.

Fig. 9 shows our low-temperature resistivity $\rho(T)$ for the ferromagnetic t=0.975 crystal. It could be fit down to 3 K to the expression

$$\rho(T)=\rho_0+AT^2 \qquad (2)$$

where $\rho_0=4.2\times 10^{-4}$ $\Omega cm$ and $A=6.2\times 10^{-8}$ $\Omega cm/K^2$; it was not necessary to add a contribution to the electron scattering by a soft optical mode. Above 100 K, $\rho(T)$ deviates from Eq. (2) as it approaches a linear temperature dependence. Since the t=0.975 crystal behaves as a typical De Gennes ferromagnetic metal, we attribute the $T^2$ dependence to a combination of electron-electron and electron-magnon scattering in a single-magnon process. In this sample, there was no evidence of the two-magnon processes predicted by Kubo and Ohata[21] for a half-metallic ferromagnet ($\rho\propto T^{4.5}$), nor the other unconventional scattering process in nearly half-metallic ferromagnets ($\rho\propto T^3$, $T^{2.5}$).[22] We attempted to stabilize the unique FM phase on traversing from one Type-CE phase to the other of Fig. 15 by applying hydrostatic pressure to the t=0.973 crystal, $Pr_{0.5}(Ca_{0.8}Sr_{0.1})_{0.5}MnO_3$. Where there is a transition from localized to itinerant electronic behavior, the (Mn-O) bond is more compressible than the (A-O) bond, which makes dt/dP>0.[23] This anomalous situation occurs because, from the virial theorem, the equilibrium (Mn-O) bond length for itinerant electrons is smaller than that for localized electrons. However, we have also observed that pressure stabilizes orbital ordering,[17] in which case the FM phase at $t_c$ would be suppressed by pressure if it were a vibronic phase of larger volume with orbital disorder (OD). The increase in the room-temperature volume at $t_c$, Fig. 2, indicates that pressure should suppress the room-temperature phase at $t=t_c$ to stabilize orbital order and/or a ferromagnetic metallic (FMM) phase. The uniqueness of the phase at $t=t_c$ extends to above room temperature although without suppressing formation of superparamagnetic (SP) clusters below T*.

The $\alpha(T)$ data of Fig. 10 allow us to follow the variation of $T_{CO}$, and $T_C$ with pressure; the charge-ordering temperature $T_{CO}$ was also tracked independently by following the sharp increase in $\rho(T)$ on cooling through $T_{CO}$. Fig. 11 shows the variation of $T_{CO}$ and $T_C$ with pressure, where $T_{CO}(\rho)$ is from $\rho(T)$ and $T_{CO}(\alpha)$ is from $\alpha(T)$. As can be seen in Fig. 10, $T_C$ and $T_{CO}$ cross at P≈12.8 kbar, which corresponds to a $t_c\approx 0.9753$ as calculated from the ratio $(dT_C/dt)/(dT_C/dP)$ of $(La_{1-y}Nd_y)_{0.7}Ca_{0.3}MnO_3$ (see ref. [24]). Figure 10 shows that the magnitude of the temperature-independent $\alpha(T)$ above $T_{CO}$ or $T_C$ does not change with pressure, indicating that the fraction of Zener polarons does not change significantly. To be noted is the thermal hysteresis. With increasing temperature, the more conductive ferromagnetic phase is completely suppressed by CO whereas with decreasing temperature, the ferromagnetic phase is not completely suppressed once nucleation is initiated below $T_C$; however, the volume fraction of the ferromagnetic phase decreases with increasing pressure.

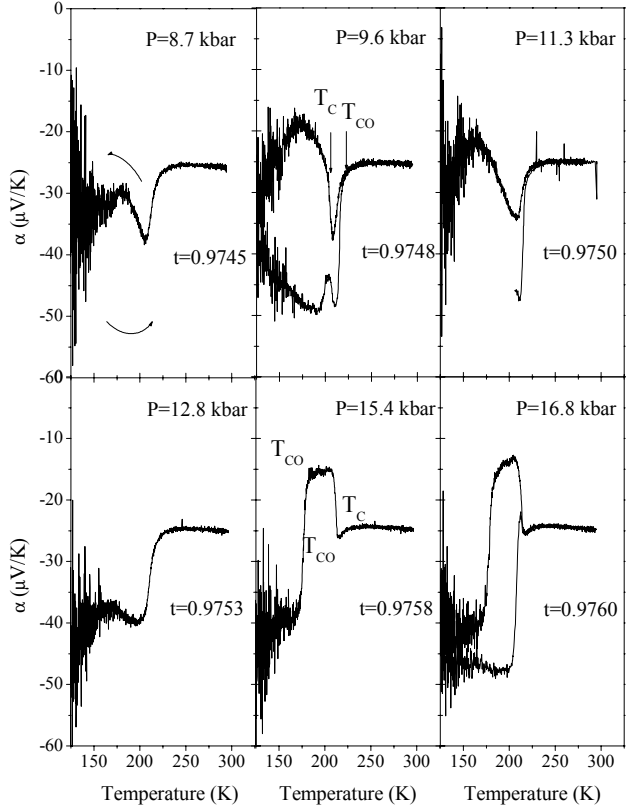

**Figure 10.-** Effect of pressure on the temperature dependence of the thermolectric power $\alpha(T)$ for $Pr_{0.5}(Ca_{1-x}Sr_x)_{0.5}MnO_3$, x=0.1; t=0.973. The meaning of the symbols and the relationship between pressure and tolerance factor is explained in the text.

Moreover, the volume fraction of ferromagnetic phase increases sharply on cooling through $T_C>T_{CO}$ at pressures P≥15.4 kbar, but the volume fraction of the ferromagnetic phase is sharply reduced below a critical temperature $T_{CO}$, and $T_{CO}$ exhibits a marked thermal hysteresis. The data are compatible with a change from a distinguishable FM minority phase having a $T_C<T_{CO}$ to a majority FM phase having a $T_C>T_{CO}$; OO at $T_{OO}=T_N$ transforms the majority phase to a CE-AF insulating phase. (see also figure 11). A FMM phase would have a shorter equilibrium (Mn-O) bond length and therefore a larger effective tolerance factor; an OO phase can compete with the FMM phase as is seen to occur in Fig. 15 for tolerance factors $t>t_c$, whereas the unique FMM phase is stable to lowest temperatures. This fact distinguishes the FMM phase at $t=t_c$ from the FMM phase at larger t even though they have similar transport properties. In Fig. 15 we distinguish a ferromagnetic vibronic FMV phase from a FMM phase where the FM phase competes with the CO phase.

A remarkable feature of the phase at $t=t_c$ is a volume larger than expected from the evolution of volume with t in Fig. 2. A larger volume implies a larger effective <A-O> bond length and a smaller mean bending angle $\phi$ of the (180°-$\phi$) Mn-O-Mn bonds, which is the equivalent of an effective tolerance factor $t_{eff}>t_c$. From Fig. 15, stabilization of a FMM phase to lowest temperatures would correspond to a $t_{eff}>0.985$. The volume contracts on cooling through $T_C$ (0.9% at 77 K) in accordance with the virial theorem for a transition from polaronic to itinerant electrons. Because of the larger room-temperature volume, the distinguishable FMM phase appearing at $t_c$ is suppressed by the application of a modest hydrostatic pressure at room temperature. We propose that where phase fluctuations occur at a small length scale, the instability of the equilibrium Mn-O bond length at the crossover from localized to itinerant electronic behavior introduces fluctuating $O^{2-}$-ion displacements perpendicular to the Mn-O-Mn bond axes that are large; these fluctuations would induce fluctuating displacements of the A cations and a strong suppression of the phonons in the paramagnetic phase. The two competing phases in this case are a CO localized-electron phase and the formation of molecular orbitals in Zener polarons in which charge differentiation is suppressed; condensation of Zener polarons into an itinerant-electron phase occurs below a Curie temperature $T_C$. Below $T_C>T_{CO}$, stabilization of itinerant $\sigma^*$ electons of e-orbital parentage can be expected to dampen the $O^{2-}$-ion fluctuations so as to induce static antiferroelectric displacements of the A-site cations while restoring the phonons. Our finding is consistent with that of Tokura *et al.*[8] who applied pressure to the conductive FM phase of $(Nd_{0.125}Sm_{0.875})_{0.5}Sr_{0.5}MnO_3$ with tolerance factor $t_c$; they reported the appearance of a static CO (or OO) second phase having a volume fraction that increased with pressure below the $T_{OO}=T_N$ transition temperature. Their pressure experiments increased t to $t>t_c$.

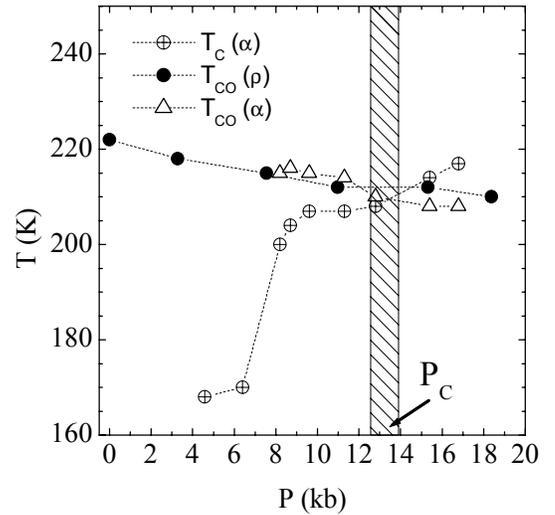

**Figure 11.-** Evolution of the critical temperatures with pressure for the crystal $Pr_{0.5}(Ca_{1-x}Sr_x)_{0.5}MnO_3$, x=0.1; t=0.973. $P_C$ represents the pressure at which $T_{CO}$ and $T_C$ cross each other. The value of the tolerance factor at this pressure roughly coincides with that at which the localised to itinerant transition takes place in the phase diagram of Figure 2 (t=0.975).

The thermal conductivities $\kappa(T)$ for $Pr_{0.5}(Ca_{1-x}Sr_x)_{0.5}MnO_3$ crystals with x=0.10, 0.15, 0.20 (corresponding to t=0.973, 0.974, 0.975) are shown in Fig. 12. Also shown is the electronic contribution of the most conductive x=0.20 (t=0.975) crystal as obtained from $\rho(T)$ and the Wiedemann-Franz ratio. The $\kappa(T)$ values of all three samples are suppressed above 250 K where fluctuating SP clusters or ferromagnetically coupled two-manganese Zener polarons suppress the phonons. The phonon contribution remains strongly suppressed down to 60 K in the x=0.10 and 0.15 samples, which is characteristic of a dynamic phase segregation by cooperative atomic displacements.[17] At t=$t_c$ (x=0.20), a step appears in $\kappa(T)$ at $T_C$ where the two-phase fluctuations change to a single FM phase. The $\rho(T)$ and $\kappa(T)$ data of the unique FMM phase are inconsistent with disorder of localized orbitals; suppression of the phase by pressure would normally be incompatible with itinerant electrons, but it is to be expected if A-cation displacements increase the effective A-cation radius. Our preliminary structural data suggest the unique ferromagnetic phase has a lower symmetry below $T_C$.

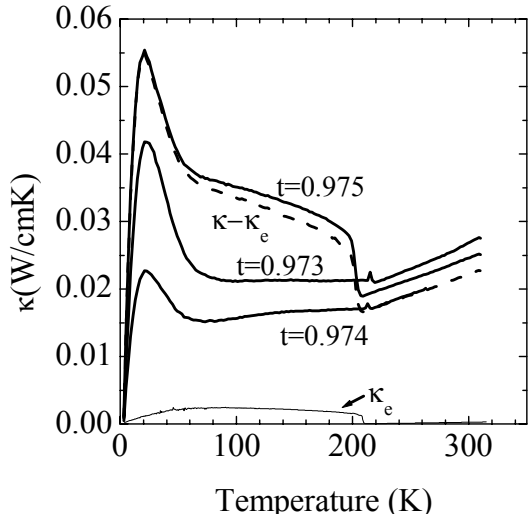

**Figure 12.-** Temperature dependence of the thermal conductivity for $Pr_{0.5}(Ca_{1-x}Sr_x)_{0.5}MnO_3$, x=0.1, x=0.15 and x=0.2. The electronic contribution to the thermal conductivity, $\kappa_e$, is also indicated for the most conductive sample (x=0.2).

Assignment of the phases and phase boundaries below $T_C$ in the interval $t_c<t<0.990$ is based on available neutron-diffraction data and our indirect probes. Itinerant electrons of the FMM phase do not allow either CO or localized-electron OO. As t decreases from 0.990, the width $W=W_\sigma exp(-\lambda\varepsilon/\hbar\omega O)$ of the $\sigma^*$ band decreases as $\omega O(\phi)$ decreases with increasing bending of the (180º-$\phi$) Mn-O-Mn bond angle.[25] As the $\sigma^*$ band narrows, the appearance of a Type-A AF phase below a $T_{OO}=T_N$ signals a change from a partially filled 3D $\sigma^*$ band to a quarter-filled 2D ($x^2$-$y^2$) $\sigma^*$ band and empty ($3z^2$-$r^2$) orbitals. De Gennes double exchange within the (001) planes is ferromagnetic, the $t^3$-O-$t^3$ superexchange between (001) planes couples them antiferromagnetically. The exact width and location of the Type-A AF phase is difficult to locate since the samples investigated by neutron diffraction[26] have shown the coexistence of the Type-A AF and FMM phases below $T_N$. We locate the Type-A AF phase in a small $\Delta t$ interval near t=0.985. As t decreases further, a charge-density wave develops within the (001) planes below $T_{OO}=T_N$ to give a Type-CE AF order. The zig-zag nature of the ferromagnetic stripes that are coupled antiparallel to one another within the (001) planes argues against 1D itinerant-electron stripes. Nevertheless the two Type-CE AF phases appearing either side of $t_c$ can be distinguished if the one for t<$t_c$ has localized e electrons at Mn(III) and ordering of Mn(IV), whereas the Type-CE AF phase at t>$t_c$ contains two-manganese Zener polarons having the e electron of a Mn(III)-O-Mn(IV) pair in a molecular orbital and no CO. An observed single valence state $Mn^{3.5+}$ for all the Mn atoms[27] in this Type-CE AF phase argues against a return to the localized-electron Type-CE AF phase found for t<$t_c$. Observation[28] of the coexistence of Type-A and Type-CE AF phases below $T_{OO}=T_N$ indicates that these phases are also separated by a two-phase interval $\Delta t$; we assume this $\Delta t$ is where $T_N$ is independent of t in Fig. 15. The $Pr_{0.5}(Ca_{0.5}Sr_{0.5})_{0.5}MnO_3$ crystal with t=0.980 is located near the boundary of the Type-CE AF phase. The $\alpha(T)$ curve of Fig. 13 provides a measure of $T_C$ and $T_{OO}=T_N$ in this crystal. On the other hand, for $Pr_{0.5}(Ca_{0.25}Sr_{0.75})_{0.5}MnO_3$, t=0.9846, (also in figure 13), the pronounced increase in $\alpha(T)$ as the temperature is reduced indicates a trapping out of the mobile polarons as the sample approaches $T_C$. Below $T_N$, the slight increase in the magnitude of the thermoelectric power reflects the absence of charge localization below this temperature as the crystal develops the Type-A AF ordering.

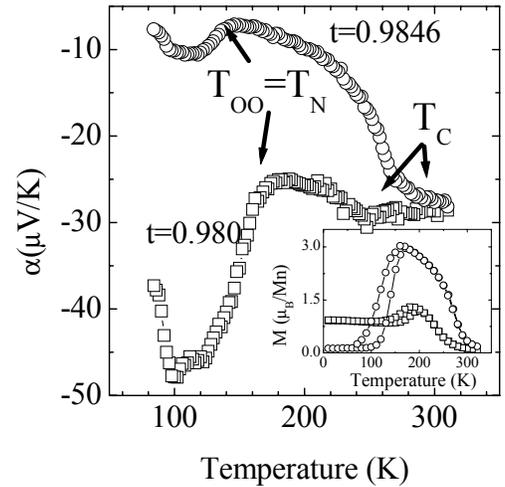

**Figure 13.-** Thermal evolution of the thermoelectric power in $Pr_{0.5}(Ca_{1-x}Sr_x)_{0.5}MnO_3$, x=0.5 and x=0.75; t=0.98 and t=0.9846 respectively. The magnetization vs. temperature (ZFC-FC, H=1 T) is shown in the inset.

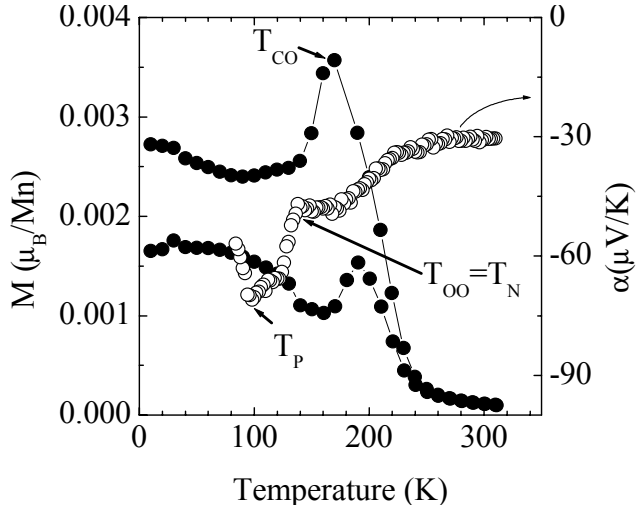

**Figure 14.-** Temperature dependence of the magnetization (●) FC-ZFC, H=10 Oe, and thermoelectric power (O) for $Pr_{0.5}(Ca_{1-x}Sr_x)_{0.5}MnO_3$, x=0.25; t=0.976.

Fig. 14 shows $\alpha(T)$ and $M(T)$ in 10 Oe for a t=0.976 crystal lying in the two-phase region between the FM and the Type-CE AF phase below $T_N$. The ZFC M(T) curve was taken in 10 Oe after cooling in zero field, the FC curve on cooling in H=10 Oe. The $\alpha(T)$ curve is to be compared with the ZFC M(T) curve. A $T_C \approx 230$ K in a ferromagnetic phase is only a little higher than the $T_{CO}$ of the paramagnetic phase. As in the t=0.973 crystal with $T_C > T_{CO}$ under a pressure $P \geq 12.4$ kbar, the paramagnetic phase undergoes CO below a $T_{CO} \geq T_{OO} = T_N \approx 140$ K, and there is a large thermal hysteresis in $T_{CO}$ just as was observed for the t=0.973 crystal under pressures $P \geq 12.4$ kbar. The data also suggest the volume fraction of the more conductive ferromagnetic phase grows below a temperature $T_p$ relative to the Type-CE AF phase. Thus pressure is seen to be equivalent to an increase of the tolerance factor t, i.e. dt/dP>0 at the crossover from localized to itinerant electronic behavior; but pressure applied at room temperature also suppresses the appearance of the unique FM phase.

It is of interest to note that the $t_c=0.975$ for the $Pr_{0.5}(Ca_{1-x}Sr_x)_{0.5}MnO_3$ system is the same as that for $Ln_{0.7}A_{0.3}MnO_3$ systems with Ln=La, Pr or Nd and A =Ca or Sr.[24,29] The critical effect of the tolerance factor on the transition from localized to itinerant electronic behavior in the system $Ln_{0.7}A_{0.3}MnO_3$ was attributed by Rivadulla et al.[30] to the critical dependence of the Jahn-Teller vibrational anisotropy with the Mn-O-Mn bond angle. On the other hand, Egami and Louca[31] suggested that this effect arises from the changes in the polaron formation energy, which strongly depends on the ionic size. Both approaches predict a sudden drop of the electron-lattice coupling constant above a particular value of the tolerance factor, which in principle should be also applicable at x=0.5. The possibility of such a first-order transition in half-doped manganites has also been predicted by computational analysis.[32]

Finally, we noted that crystals with larger variance $\sigma^2$ of the A-site cation size suppressed the $T_C$ of the unique FMM phase at $t_c$. This observation is consistent with the conclusions of Burgy et al.[7] who assigned this phenomenon to the effect of "quenched disorder". However, we find that the predicted quantum critical point is replaced by stabilization of a FMM phase inserted at the predicted crossover boundary. Stabilization of this FMM phase may be suppressed by a larger perturbation of the periodic potential in samples with a larger variance.

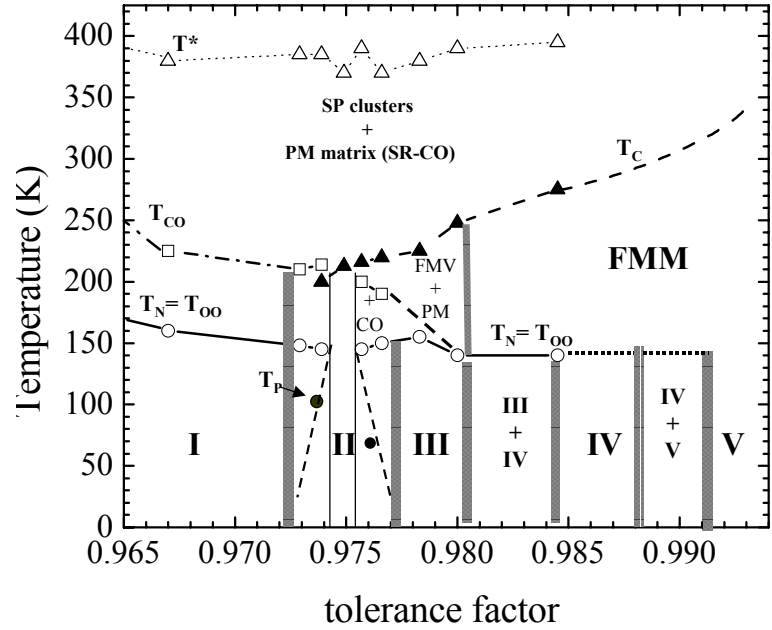

**Figure 15.-** Our phase diagram for $Ln_{0.5}A_{0.5}MnO_3$ (Ln=La, Pr, Y, Sm, etc; A=Ca, Sr) perovskites. In the paramagnetic range, above $T_C$ and $T_{CO}$, SR-CO stands for short-range CO fluctuations in the paramagnetic matrix. The different regions at lower temperatures correspond to: (I) Type-CE antiferromagnetic insulating (with CO of the Mn(III) and Mn(IV) ions), (II) ferromagnetic metallic (larger volume), (III) Type-CE antiferromagnetic insulating (Zener polarons), (IV) Type-A antiferromagnetic, (V) ferromagnetic metallic (smaller volume). Two-phase fluctuations with CO majority phase between I and II, FMM majority phase between II and III and PM minority phase for $T_{CO}<T<T_C$ becomes CO minority phase below $T_{CO}$ for t>0.975. Shaded boundaries are estimates. We identify $T_{OO}$ stands for long-range orbital ordering (see text).

## IV.- CONCLUSIONS

Single-crystal measurements of the physical properties of $Ln_{0.5}A_{0.5}MnO_3$ crystals, where Ln is one or a mixture of two lanthanide ions and A is one or a mixture of two alkaline earths, have revealed a critical tolerance factor $t_c \approx 0.975$ within the crossover range from localized-electron behavior for $t<t_c$ to itinerant-electron behavior for $t>t_c$.

With $t<t_c-\Delta t_1$ in the crystal system studied, long-range CO occurs below a $T_{CO}$ and long-range OO occurs below a $T_{OO}$ that coincides with AF ordering below a $T_N=T_{OO}$. Short-range OO in the interval $T_N<T<T_{CO}$ suppresses magnetic order. The CO and OO antiferromagnetic phase has Type-CE magnetic order. With $t>t_c+\Delta t_2$, charge ordering is suppressed and a FMM phase is stabilized below a Curie temperature $T_C$. As t is lowered toward $t_c$, the σ* band of e-orbital parentage is narrowed and there is a change from a 3D σ* band to a 2D σ* band to give Type-A AF order with ferromagnetic (001) planes coupled antiparallel below a $T_N=T_{OO}$. As t is lowered further, this phase is followed by the appearance of Type-CE AF order in which the ferromagnetic (001) planes break up into ferromagnetic zig-zag chains coupled antiparallel to one another; the chains appear to contain two-manganese Zener polarons in which the average manganese valence is 3.5+.[27,33] The charge-ordered Type-CE AF phase at $t<t_c$ contains distinguishable Mn(IV) and Mn(III) ions. At $t=t_c$, a unique ferromagnetic phase suppresses both CO and OO down to lowest temperatures. A larger volume of the room-temperature phase at $t=t_c$ made it impossible to obtain the unique FMM phase by applying pressure to a $t<t_c$ crystal. Although a dt/dP>0 caused t to increase with P as anticipated at a crossover from localized to itinerant electronic behavior, the Type-CE-AF + FM two-phase region for $t<t_c$ changes continuously under pressure as $T_C$ of the FM phase crosses $T_{CO}$ of the localized-electron phase.

All these results are summarized in the phase diagram of Fig. 15.

## ACKNOWLEDGMENTS

The authors thank the NSF, the Robert A. Welch Foundation, Houston, TX, and the TCSUH of Houston, TX. F. R. wants to thank the Fulbright Foundation and MECD (Spain) for a postdoctoral fellowship.